\documentclass{CIRED-v1}
\usepackage{euscript}
\usepackage{amsfonts}
\usepackage[colorlinks]{hyperref}       
\usepackage{url}            
\usepackage{amsmath}
\usepackage{amssymb}
\usepackage{graphicx}
\usepackage{verbatim}
\usepackage{subcaption}
\usepackage{tabularx,booktabs}
\usepackage[dvipsnames]{xcolor}
\usepackage{colortbl}
\usepackage{bm}
\usepackage{comment}
\usepackage{caption}
\usepackage{threeparttable}
\usepackage{wrapfig}
\usepackage[framemethod=tikz]{mdframed}
\usetikzlibrary{arrows,intersections,chains,calc}
\usetikzlibrary{positioning}
\usetikzlibrary{shapes}

\usepackage{pgfplots}
\pgfplotsset{compat=newest}
\usepackage{eurosym}

\usepackage{soul}

\begin{document}

\title{A Linear Model for Distributed Flexibility Markets and DLMPs: A Comparison with the SOCP Formulation}
\author{Anibal Sanjab\ad{1\corr}, Yuting Mou\ad{1}, Ana Virag \ad{1}, Kris Kessels \ad{1}}
\address{\add{1}{Unit Energy Technology, Flemish Institute for Technological Research (VITO) and EnergyVille, Genk, Belgium}
\email{Corresponding author: anibal.sanjab@vito.be}\vspace{-0.4cm}}

\begin{abstract}
This paper examines the performance trade-offs between an introduced linear flexibility market model for congestion management and a benchmark second-order cone programming (SOCP) formulation. 
The linear market model incorporates voltage magnitudes and reactive powers, while providing a simpler formulation than the SOCP model, which enables its practical implementation. The paper provides a structured comparison of the two formulations relying on developed deterministic and statistical Monte Carlo case analyses using two distribution test systems (the Matpower 69-bus and 141-bus systems). The case analyses show that with the increasing spread of offered flexibility throughout the system, the linear formulation increasingly preserves the reliability of the computed system variables as compared to the SOCP formulation, while more lenient imposed voltage limits can improve the approximation of prices and power flows at the expense of a less accurate computation of voltage magnitudes. 
\end{abstract}

\maketitle

\section{Introduction}
With\footnote{This paper has been accepted and presented at the CIRED'21 conference. The paper in its final version is published in the proceedings of the conference.} the proliferation of distributed renewable energy resources and increasing electrification of the consumer energy space, a higher stress is projected to be placed on the operation of the distribution system. This is stemming from the increasing risk of surpassing distribution equipment capacity, voltage ratings, and line flow constraints. Alleviating such constraints, commonly referred to as congestion management, can be achieved by these very same generation and load resources by leveraging their flexibility, i.e., the temporary adjustment of their generation and consumption patterns as a service to the distribution system. In addition, the value of this distributed flexibility can even extend beyond the limits of the distribution system by providing ancillary services to the transmission system. Indeed, leveraging distributed flexibility, at different voltage levels, is a main component for achieving the intended renewable energy integration and climate goals~\cite{Ceer}. 

Different methods for leveraging this distributed flexibility have recently emerged, where the recommendation for market-based mechanisms has been increasingly growing~\cite{Ceer}. However, the offering or purchasing of such flexibility resources on the distribution level requires a market design that i) accounts for the physical limits of the distribution system, ensuring its secure operation, ii) provides a pricing mechanism which allows a competitive and fair remuneration for flexibility service providers and purchasers (while minimizing the possibilities of exerting market power), and iii) requires manageable computational complexity enabling a reliable, transparent, and timely market clearing. Along these lines, market models for distribution systems based on a second order cone programming (SOCP) formulation, and adopting distribution locational marginal prices (DLMPs), have been recently proposed in the literature~\cite{DLMP_SOCP,DLMP_SOCP_HLC}. The premise of the SOCP formulation is that it provides a convex formulation, which closely approximates the non-convex AC optimal power flow model. However, the numerical expensiveness of the SOCP solution can still pose challenges for implementation in practice, especially in markets with thousands of nodes, bids, and constraints that must be cleared in a time-restrictive manner~\cite{SmartNet}. For example, in the context of the Horizon 2020 CoordiNet project, all of the demonstration campaigns focusing on congestion management have opted -- beyond the pre-qualification phase -- for linear market representations (using, e.g., derived sensitivity factors) when considering network models in the market formulation\footnote{The different market designs in the CoordiNet project are discussed in D6.2, D2.1, and D1.3 reports available at \url{coordinet-project.eu/publications/deliverables}.}. 

This paper, on the other hand, considers a linear flexibility market model for congestion management in distribution systems, aiming at assessing its accuracy as compared to the SOCP formulation. The linear market design relies on a linear distribution network representation and a linearization of flow constraints. The resulting linear programming (LP) formulation incorporates voltage magnitudes and reactive powers while providing a simpler market clearing and computation of DLMPs as compared to the SOCP formulation, improving its potential for practical implementation. For benchmarking the linear formulation, the paper develops a structured case analysis using a deterministic and Monte Carlo-based statistical comparison between the results generated using the LP and SOCP formulations. This comparison i) assesses the quality of the generated results of the LP formulation (considering DLMPs, power flows, voltage magnitudes, and participants' revenues) compared to the SOCP, and ii) allows the identification of the key elements driving the convergence of the outcomes of the two formulations. In this regard, our case analyses show that with the increase of the spread and availability of flexibility over different distribution nodes (a condition that is increasingly expected in emerging distribution flexibility markets), the LP formulation generates results that closely capture those of the SOCP formulation. In addition, the case analyses highlight the effect of strictness of voltage limits on the quality of the results of the LP formulation.  


The focus of the paper is, hence, on identifying any accuracy trade-offs between the LP and SOCP formulations. As these trade-offs may vary between different systems, our analyses focus on specific case studies using the Matpower 69-bus and 141-bus distribution test systems~\cite{Matpower}. 

\section{Flexibility Market Model}
Consider a radial distribution system composed of $\mathcal{N}$ nodes (where $n^o \in \mathcal{N}$ denotes the root node) 
and $\mathcal{L}$ distribution lines forming a graph (tree), $\mathcal{G}(\mathcal{N},\mathcal{L})$. We let $A(i)$ denote the ancestor node of $i\in\mathcal{N}\setminus\{n^o\}$ and $\mathcal{K}(i)$ the set of predecessor nodes of $i$. For the root node $n^o$, $A(n^o)$ denotes the interconnection node with the upper-level grid (e.g., the transmission system). We let, a) $s_i=p_i+jq_i$ be the net complex power injection at bus $i$, b) $P_{A(i)i}$ and $Q_{A(i)i}$ be, respectively, the real and reactive power flowing over the line connecting $A(i)$ and $i$, measured at the sending node ${A}(i)$, c) $l_{A(i)i}$ be the magnitude squared of the current flowing over $\{A(i),i\}\in\mathcal{L}$, and d) $v_i$ be the magnitude squared of the voltage at node $i\in\mathcal{N}$. Here, $P_{A(n^o)n^o}$ and $Q_{A(n^o)n^o}$ denote the interface flows with the upper-level grid.
As for the line parameters, we let $r_{A(i),i}$ and $x_{A(i),i}$ be, respectively, the resistance and reactance of line $\{A(i),i\}\in\mathcal{L}$.   

Consider a distribution system operator (DSO) aiming to purchase flexibility to alleviate congestion constraints (i.e., surpassing of line flow or nodal voltage magnitude limits) that are expected to occur at the base generation and load profiles given, respectively, by vectors $\boldsymbol{p}^o$ and $\boldsymbol{d}^o$, where $p^o_i$ and $d^o_i$ denote the base generation and load at node $i\in\mathcal{N}$. 
Let $\Delta p_i^{+}$ and $\Delta p_i^{-}$ correspond, respectively, to the volumes of increase and reduction in generation (respectively, upward and downward flexibility of a generation resource) connected at the distribution node $i$. In addition, we let $\Delta d_i^{+}$ and $\Delta d_i^{-}$ represent upward and downward flexibility provided by the demand at distribution node $i$ (i.e., the volumes of reduction and increase in demand at that node). We  let $c_{p_i}^{+}$, $c_{p_i}^{-}$, $c_{d_i}^{+}$, and $c_{d_i}^{-}$ represent the cost offered by adjustable generation and loads on the distribution level for the activation of, respectively, $\Delta p_i^{+}$, $\Delta p_i^{-}$, $\Delta d_i^{+}$, and $\Delta d_i^{-}$. %

As the DSO's goal is to resolve congestion issues at minimum cost, we formulate it's flexibility market clearing as an LP problem as follows.\vspace{-0.4cm}

\begin{align}
\min_{\Delta \boldsymbol{p},\Delta \boldsymbol{d}}\,\,\,\sum\limits_{i=1}^{N}\left(c_{p_i}^{+}\Delta p_i^{+} - c_{p_i}^{-}\Delta p_i^{-} + c_{d_i}^{+}\Delta d_i^{+} - c_{d_i}^{-}\Delta d_i^{-}\right), \label{eq:ObjectiveCM_onlyD}
\end{align}\vspace{-0.3cm}
Subject to:
\begin{align}
p_i\!=\!p^{o}_i\!+\!\Delta p_i^{+}\!-\!\Delta p_i^{-}\!-\!d^{o}_i \!+\! \Delta d_i^{+} \!-\! \Delta d_i^{-}, \forall i\in \mathcal{N},  \label{eq:AdjustedInjection_OnlyD_Lin1} 
\end{align}\vspace{-0.8cm}
\begin{align}
p_i+P_{A(i)i}-\sum\limits_{k\in\mathcal{K}(i)}P_{ik}=0,\,:(\lambda_i^{L}),\,\forall i\in\mathcal{N}, \label{eq:LinDistFlow1_OnlyD_Lin1}\\
q_i+Q_{A(i)i}-\sum\limits_{k\in\mathcal{K}(i)}Q_{ik}=0,\,\forall i\in\mathcal{N},\label{eq:LinDistFlow2_OnlyD_Lin}
\end{align}\vspace{-0.8cm}
\begin{align}
v_i=v_{\mathcal{A}(i)}-2r_{A(i)i}P_{A(i)i}-2x_{A(i)i}Q_{A(i)i},\,\forall i\in\mathcal{N},\label{eq:LinDistFlow3_DSOLocal_OnlyD_Lin1}
\end{align}\vspace{-0.8cm}
\begin{align}
\alpha_m P_{A(i)i} \!+\! \beta_m Q_{A(i)i} \!+\! \delta_m  S_{A(i)i}^{\textrm{max}} \!\leq\! 0, \forall m \!\in\! \mathcal{M}, \{A(i),i\} \!\in\!\mathcal{L}, \label{eq:FlowLimit_OnlyD_Lin1}
\end{align}\vspace{-0.8cm}
\begin{align} \label{eq:VMaxCM_onlyD_Lin1}
 v_i^{\textrm{min}} ~\!\leq\! v_i \!\leq\! v_i^{\textrm{max}}, \,q_i^{\textrm{min}} \!\leq\! q_i \!\leq\! q_i^{\textrm{max}},\,\, \forall i\in\mathcal{N},
\end{align}
\vspace{-0.8cm}
\begin{align}\label{eq:GenBidLimCMp_onlyD1}
0\!\leqslant\! \Delta p_i^{+} \!\leqslant\! \Delta p_i^{+,\textrm{max}},\,0\!\leqslant\! \Delta p_i^{-} \!\leqslant\! \Delta p_i^{-,\textrm{max}}, \, \forall i\in\mathcal{N}, 
\end{align}
\vspace{-0.8cm}
\begin{align} \label{eq:DimBidLimCMp_onlyD1}
0\!\leqslant\! \Delta d_i^{+} \!\leqslant\! \Delta d_i^{+,\textrm{max}}, \,0\!\leqslant\! \Delta d_i^{-} \!\leqslant\! \Delta d_i^{-,\textrm{max}},\, \forall i\in\mathcal{N}.
\end{align}
\vspace{-0.4cm}

Here, $\Delta \boldsymbol{p}$ and $\Delta \boldsymbol{d}$ are vectors of decision variables (i.e. $\Delta p_i^{+}, \,\Delta p_i^{-},\, \Delta d_i^{+}$ and $\Delta d_i^{-}$) at each $i\in\mathcal{N}$.  
(\ref{eq:AdjustedInjection_OnlyD_Lin1}) returns the aggregated net injection $p_i$ at node $i$ taking into consideration the baseline generation and load profiles at this node as well as the activated flexibility.  
The equality constraints in (\ref{eq:LinDistFlow1_OnlyD_Lin1})-(\ref{eq:LinDistFlow3_DSOLocal_OnlyD_Lin1}) capture the linearized power flow equations in radial distribution networks following directly from the \emph{LinDistFlow} formulation~\cite{LinDistFlow}. The \emph{LinDistFlow} formulation approximates the common \emph{relaxed branch flow} equations~\cite{LinDistFlow} (known as \emph{DistFlow}) by dropping the branch loss and shunt components.   
We denote the Lagrange multiplier of the nodal power balance constraint (\ref{eq:LinDistFlow1_OnlyD_Lin1}) by $\lambda_i^{L}$, which constitutes the DLMP for this linear formulation at node $i$.
Constrain~(\ref{eq:FlowLimit_OnlyD_Lin1}) is a linearization of the complex flow limit constraint typically represented in SOCP formulations as $(P_{A(i)i})^2 + (Q_{A(i)i})^2 \leq (S_{A(i)i}^{\textrm{max}})^2$~\cite{DLMP_SOCP}, where $S_{A(i)i}^{\textrm{max}}$ is the maximum flow capacity of line $\{A(i),i\}$. 
This linearization, as proposed in~\cite{DGHostingCapacity}, is a polygonal inner-approximation which transforms the feasibility region of the flow limit constrain into a polygon whose number of edges are given by the size of the approximation set $\mathcal{M}$, denoted by $M$. The values of $\alpha_m$, $\beta_m$, and $\gamma_m$ define this polygon such that all vertices of the polygon would lie on the circle of radius $S_{A(i)i}^{\textrm{max}}$ (a more detailed explanation of this approximation is provided in~\cite{DGHostingCapacity}). 
The constraints in~(\ref{eq:VMaxCM_onlyD_Lin1}) limit the voltage magnitude and reactive power injections at the different buses to predefined lower and upper limits to ensure operational stability and preserve real-reactive power operational and capacity limits of load and generation.  
Constraints (\ref{eq:GenBidLimCMp_onlyD1}) and~(\ref{eq:DimBidLimCMp_onlyD1}) capture the limits of the submitted bids.  

The counterpart non-linear formulation to the presented linear formulation in (\ref{eq:ObjectiveCM_onlyD})-(\ref{eq:DimBidLimCMp_onlyD1}) is the SOCP formulation (such as the one in~\cite{DLMP_SOCP}), which, instead of relying on the \emph{LinDistFlow} power flow equations, makes use of the relaxed \emph{branch flow} equations~\cite{DLMP_SOCP} and uses a conic feasibility space in the definition of the constraints. The SOCP formulation is, hence, similar to the formulation in (\ref{eq:ObjectiveCM_onlyD})-(\ref{eq:DimBidLimCMp_onlyD1}), but in which constraints (\ref{eq:LinDistFlow1_OnlyD_Lin1})-(\ref{eq:FlowLimit_OnlyD_Lin1}) are replaced by:

\begin{align}
p_i\!+\!(P_{A(i)i}\!-\!r_{A(i)i}l_{A(i)i})\!-\!\!\!\!\sum\limits_{k\in\mathcal{K}(i)}P_{ik}\!-\!g_i^sv_i\!=\!0,\,:(\lambda_i^{S}),\,\forall i\!\in\!\mathcal{N},\label{eq:DistFlow1}
\end{align}\vspace{-0.8cm}
\begin{align}
q_i\!+\!(Q_{A(i)i}\!-\!x_{A(i)i}l_{A(i)i})\!-\!\!\!\!\sum\limits_{k\in\mathcal{K}(i)}Q_{ik}\!-\!b_i^sv_i\!=\!0,\forall i\in\mathcal{N},\label{eq:DistFlow2}
\end{align}\vspace{-0.8cm}
\begin{flalign}
v_i\!=\!v_{\mathcal{A}(i)}\!&-\!2(r_{\mathcal{A}(i)i}P_{\mathcal{A}(i)i}\!+\!x_{\mathcal{A}(i)i}Q_{\mathcal{A}(i)i})&\!\nonumber\\
&+\!(r_{\mathcal{A}(i)i}^2\!+\!x_{\mathcal{A}(i)i}^2)l_{\mathcal{A}(i)i},\,\,\,\forall i\in\mathcal{N},\label{eq:DistFlow3}&
\end{flalign} \vspace{-0.8cm}
\begin{align}
\frac{(P_{\mathcal{A}(i)i})^2+(Q_{\mathcal{A}(i)i})^2}{l_{\mathcal{A}(i)i}}\leq v_{A(i)}, \,\forall \{A(i),i\}\in\mathcal{L},\label{eq:CurrentLimitSOCP}
\end{align}\vspace{-0.8cm}
\begin{align}
(P_{\mathcal{A}(i)i})^2 + (Q_{\mathcal{A}(i)i})^2 \leq (S_{\mathcal{A}(i)i}^{\textrm{max}})^2, \,\forall \{A(i),i\}\in\mathcal{L},\label{eq:FlowLimitSOCP}
\end{align}
where (\ref{eq:DistFlow1})-(\ref{eq:DistFlow3}) are the relaxed \emph{branch flow} equations, while (\ref{eq:CurrentLimitSOCP}) and (\ref{eq:FlowLimitSOCP}) are the added non-linear constraints. In this formulation, $y_i^s=g_i^s-jb_i^s$ is the shunt admittance from $i$ to ground, and $\lambda_i^S$ is the dual multiplier of the power balance constraint~(\ref{eq:DistFlow1}) at node $i$, corresponding to its DLMP for the SOCP formulation~\cite{DLMP_SOCP}. 


\section{Deterministic comparison}\label{sec:DeterministicComparison} 
For comparing the LP and SOCP formulations, we analyze simulation results obtained using the 141-bus Matpower test system~\cite{Matpower}, and then further corroborate the conclusions using a case analysis on the Matpower 69-bus test system~\cite{Matpower}. We slightly modify each test system by adding base supply to nodes where base demand exists, in order to represent distributed -- generation and load -- flexibility resources. The quantity of the base supply is chosen randomly following a uniform distribution spanning the range of 10\% - 90\% of the base demand provided in the test case~\cite{Matpower}. For the generation of flexibility bids, the submitted costs are drawn from a uniform distribution in the range of $[35,\,45]$ \euro/MWh for demand bids and $[45,\,55]$ \euro/MWh for supply offers. This is intended to generate a random set of offers without included biases from any strategic behavior. 
The submitted quantity of each bid is generated according to the quantity of base demand/supply on each node. More specifically, the base demand can decrease to 0 to provide an upward flexibility offer (i.e., $\Delta d_i^+$) or increase by 50\% to provide a downward bid (i.e., $\Delta d_i^-$), whereas the base supply can decrease to 0 to provide a downward flexibility bid (i.e., $\Delta p_i^-$) or increase by 50\% for an upward flexibility offer (i.e., $\Delta p_i^+$).

A fundamental difference between the proposed LP and SOCP formulations (besides how the flow and current constraints are defined) is that the LP formulation neglects branch loss terms, through the \emph{LinDistFlow} model, as compared to the loss-inclusive \emph{branch flow} model of the SOCP formulation. Hence, when the effects of the losses diminish, it is expected that the LP formulation results would tend to more closely reflect the SOCP outcomes. The effects of branch losses could be typically mitigated by shorter circuit distances (i.e. a shorter separation between generation sources and load sinks). To emulate this effect, we consider different spread levels of distributed flexibility throughout the system. A higher spread level indicates a higher availability of flexibility (i.e., flexible generation or load) throughout the system. 
When the spread increases, the distances between flexibility generation and load would be expected to decrease, hence improving the performance of the LP formulation as compared to the SOCP model.  

To capture this effect, we consider two spread levels (SLs) defined as follows: 1) Spread level 1 (SL1): flexibility bids are only offered at the end nodes of selected long laterals (where generation or load exists), and 2) Spread level 2 (SL2): flexibility bids are offered from  every node where load or generation exists.  
We next compute the solutions for the LP and SOCP formulations and compare the generated results for each of the SLs, first focusing on the 141-bus test system.\vspace{-0.4cm} 

\begin{figure}[h]
    \centering
	\begin{subfigure}{0.4\textwidth}
		\centering
		\resizebox{\textwidth}{!}{
			\includegraphics[width=7cm]{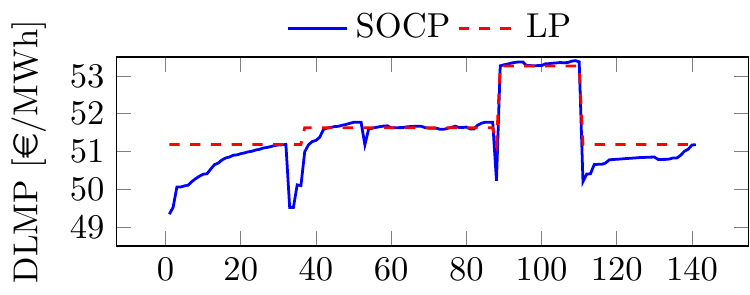}
		}
		\caption{Spread level 1}
		\label{fig:DLMP_comparison_end}
	\end{subfigure}
	\begin{subfigure}{0.4\textwidth}
		\centering
		\resizebox{\textwidth}{!}{
			\includegraphics[width=7cm]{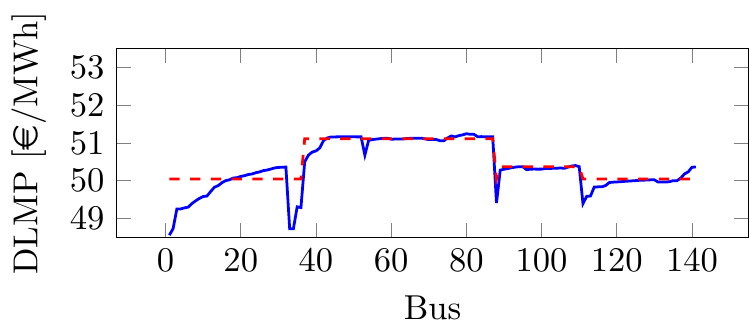}
		}
		\caption{Spread level 2}
		\label{fig:DLMP_comparison_all}
	\end{subfigure}
	\caption{DLMPs at all the buses of the 141-bus test system under different spread levels (SL1 and SL2).}
	\label{fig:DLMP_comparison}
\end{figure}\vspace{-0.4cm}

The DLMPs at the different buses for the two SLs in the 141-bus test system are shown in Fig.~\ref{fig:DLMP_comparison}. In the LP cases of both Fig.~\ref{fig:DLMP_comparison_end} and Fig.~\ref{fig:DLMP_comparison_all}, the DLMPs show three distinct values due to occurrence of congestion at two distribution lines. The DLMPs from the SOCP formulation show different values on different buses, as, in addition to line congestion and voltage limits, losses are modeled explicitly and play a key role in the formation of the DLMPs (a detailed analysis of the components of the DLMPs in the SOCP formulation is presented in \cite{DLMP_SOCP}). As shown in Fig.~\ref{fig:DLMP_comparison}, the DLMPs from the LP formulation are able to approximate those from the SOCP formulation in a relatively accurate way. Indeed, the maximum variation of DLMPs between the LP and SOCP formulation occurs at the root node and accounts for a difference between the LP result and the SOCP result of only 3.74\% and 3.06\% in, respectively, SL1 and SL2. 

In fact, in Table \ref{tab:rmse}, we compute the resulting root-mean-squared error (RMSE) of the LP market outcomes as compared to those of the SOCP, considering different system variables (namely, DLMPs, voltage magnitudes, apparent power flows, and participants' revenues), under different SLs (SL1 and SL2). For ease of comparison, the results are normalized with respect to those of SL1 except for the last row (labeled "SL2-N"), which shows the raw RMSEs for SL2. The row labeled SL2-s2 will be introduced shortly. 
As can be seen in Table \ref{tab:rmse}, the RMSEs of the DLMPs under the LP formulation as compared to the SOCP results are minimal, validating the results of Fig.~\ref{fig:DLMP_comparison}. In addition, in the case of a high spread of flexibility bids (SL2), the price differences between the LP formulation and SOCP model are smaller than in the case of SL 1, as can be also observed by the results of Fig.~\ref{fig:DLMP_comparison}.

\begin{figure}[h]
		\centering
	\begin{subfigure}{0.4\textwidth}
		\centering
		\resizebox{\textwidth}{!}{
			\includegraphics[width=7cm]{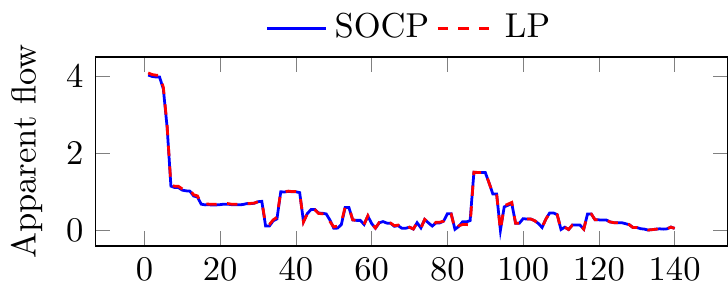}
		}
		\caption{Spread level 1}
		\label{fig:ApparentFlow_comparison_end}
	\end{subfigure}
	\begin{subfigure}{0.4\textwidth}
		\centering
		\resizebox{\textwidth}{!}{
			\includegraphics[width=7cm]{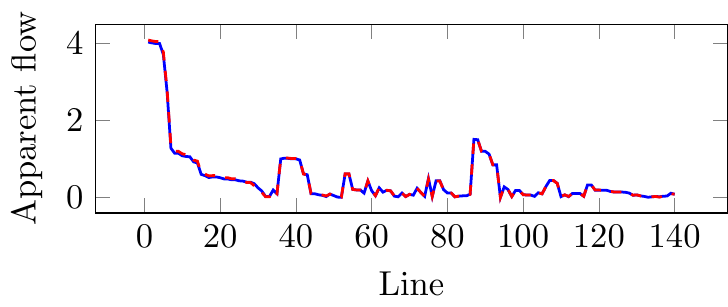}
		}
		\caption{Spread level 2}
		\label{fig:ApparentFlow_comparison_all}
	\end{subfigure}
	\caption{Apparent power flow [MVA] over the different lines within the 141-bus test system, considering different SLs.}
	\label{fig:ApparentFlow_comparison}
\end{figure}

As the key goal of the market formulation is to achieve congestion management, the performance of the LP formulation in approximating the SOCP formulation in terms of apparent power flow is of key importance. In Fig. \ref{fig:ApparentFlow_comparison}, we present the comparison between the apparent power flows resulting from the LP and SOCP formulations for both spread levels. As can be seen in Fig. \ref{fig:ApparentFlow_comparison}, the results of the LP and SOCP formulations are extremely close (manifested by coinciding flow levels at all of the branches), which motivates the practical application of the LP formulation for congestion management. The accurate computation of the apparent power flow under the LP formulation is also captured in the resulting low RMSE under the "Flow" column of Table \ref{tab:rmse}.\vspace{-0.4cm}

\vspace{-0.6cm}

\begin{table}[h]
	\caption{RMSEs of main market outcomes for different SLs \\ using the 141-bus test system (normalized with respect to \\SL1, except for SL2-N).}
	\begin{center}
		\begin{tabular}{l|c|c|c|c}\toprule
			&DLMP &Voltage &Flow&Revenue \\ \midrule
			SL1& 1 & 1 & 1& 1 \\
			SL2  & 0.70 & 0.51 & 0.92& 0.18  \\
			SL2-s2  & 0.49 & 3.73 & 0.61&0.17  \\\hline
			SL2-N  & 0.03 & 3$\times$10$^{-6}$ &0.002 & 0.004 \\\botrule
		\end{tabular}
		\label{tab:rmse}
	\end{center}
\end{table}\vspace{-0.5cm}

In addition to the spread level, the imposed voltage upper and lower limits ($v_i^\textrm{max}$ and $v_i^\textrm{min}$ in~(\ref{eq:VMaxCM_onlyD_Lin1})) also play a significant role in allowing the results of the LP formulation to converge to those of the SOCP model. Indeed, this could be expected by comparing the computation of nodal voltages under the two formulations in, respectively,~(\ref{eq:LinDistFlow3_DSOLocal_OnlyD_Lin1}) and~(\ref{eq:DistFlow3}). In the simulations of SL1 and SL2, shown in Fig.~\ref{fig:DLMP_comparison} and \ref{fig:ApparentFlow_comparison}, the voltage has been limited to $[0.99,\ 1.01]$ p.u. at all buses. To evaluate the impact of voltage limits, we create a new case labeled SL2-s2, where the voltage is restricted to $[0.9, \ 1.1]$ p.u. under SL2. The resulting RMSEs for the SL2-s2 case are shown in Table~\ref{tab:rmse}. 


By contrasting SL1 with SL2 in~Table~\ref{tab:rmse}, we can see that the LP formulation approximates the SOCP formulation more closely when the flexibility bids' spread increases. When we relax the voltage limits in SL2-s2, we can observe that the voltage differences increase compared with SL2, but, in return, better results for the DLMPs, the apparent power flows, and revenues are obtained. This shows a possible trade-off in decreasing the quality of the voltage magnitudes' computation for increasing those of the DLMPs and the power flows. The RMSEs in Table~\ref{tab:rmse}, in general, capture the minimal error introduced by the LP formulation when compared to the SOCP formulation for the different market outcomes for the 141-bus test case, attesting to the quality of the obtained results.

To further validate the results obtained from the 141-test system, we follow a similar simulation mechanism using the 69-bus test system, considering SL1 and SL2 as well. The RMSEs comparing the different outcomes obtained under the LP and SOCP formulations are shown in Table \ref{tab:rmse_69}. Table \ref{tab:rmse_69} confirms the ability of the LP formulation to reliably compute market outcomes as indicated by the minimal resulting RMSEs. In addition, Table \ref{tab:rmse_69} also showcases the effect of the increasing spread level on the quality of the obtained results corroborating the observations derived from the 141-test system.\vspace{-0.3cm}

\begin{table}[h]
	\caption{RMSEs of main market outcomes for different SLs 
	\\using the 69-bus test system (normalized with respect to SL1, \\
	except for SL2-N).}
	\begin{center}
		\begin{tabular}{l|c|c|c|c}\toprule
			&DLMP &Voltage &Flow&Revenue \\ \midrule
			SL1& 1 & 1 & 1& 1 \\
			SL2  & 0.022 & 0.24 & 0.91& 0.41  \\\botrule
			SL2-N  & 0.02 & 3$\times$10$^{-5}$ &0.001 & 0.01 \\\botrule
		\end{tabular}
		\label{tab:rmse_69}
	\end{center}
\end{table}\vspace{-0.8cm}


\section{Statistical analysis}

\begin{figure}[t!]
		\centering

	\begin{subfigure}{0.4\textwidth}
		\centering
		
		\resizebox{\textwidth}{!}{
			\includegraphics[width=7cm]{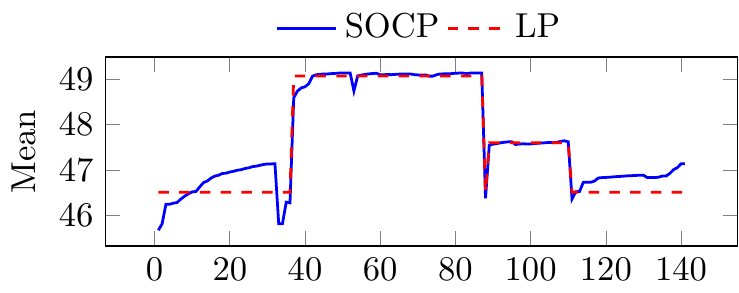}
		}
		\label{fig:DLMP_mean}
	\end{subfigure}\vspace{-0.25cm}
	\begin{subfigure}{0.4\textwidth}
		\centering
		\resizebox{\textwidth}{!}{
			\includegraphics[width=7cm]{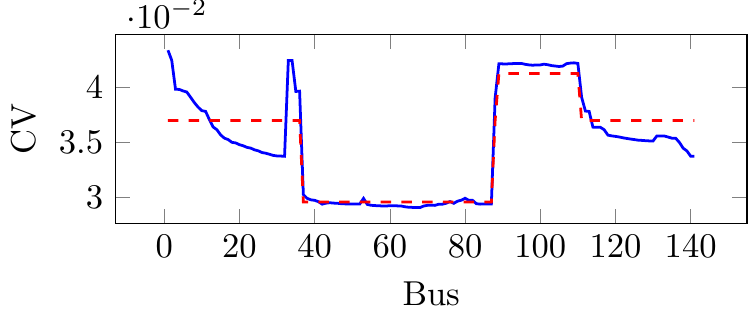}
		}
		\label{fig:DLMP_coeff}
	\end{subfigure}\vspace{-0.5cm}
	\caption{The numerical mean [\euro/MWh] and CV of the DLMPs at the different buses under the LP and SOCP formulations.}
	\label{fig:DLMP}
\end{figure}

\begin{figure}[t!]
		\centering

	\begin{subfigure}{0.4\textwidth}
		\centering
		
		\resizebox{\textwidth}{!}{
			\includegraphics[width=7cm]{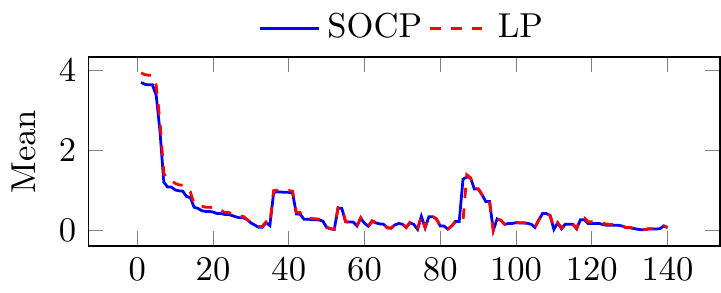}
		}
		\label{fig:ApparentFlow_mean}
	\end{subfigure}\vspace{-0.25cm}
	\begin{subfigure}{0.4\textwidth}
		\centering
		\resizebox{\textwidth}{!}{
			\includegraphics[width=7cm]{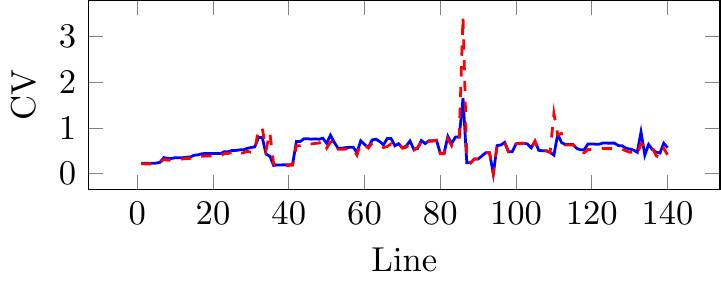}
		}
		\label{fig:ApparentFlow_coeff}
	\end{subfigure}\vspace{-0.5cm}
	\caption{The numerical mean [MVA] and CV of apparent flows over the different lines for the LP and SOCP formulations.}
	\label{fig:ApparentFlow}
\end{figure}

The results derived in Section~\ref{sec:DeterministicComparison} were based on a deterministic comparison of the outcomes of the LP and SOCP formulations. To obtain a more in-depth comparison of the results, we perform here a statistical comparison of the outcomes of the LP and SOCP formulations considering Monte Carlo~\cite{MSThesis} simulations on the 141-bus test system under the SL2 case. The random inputs in this analysis are considered to be the submitted bids. As such, instead of considering deterministic costs, we allow the cost of each bid in (\ref{eq:ObjectiveCM_onlyD}), i.e., $c_{\beta_i}^\alpha$, where $\beta\in\{p,d\}$ and $\alpha\in\{+,-\}$, to be scaled up/down by a factor of $f_{\beta_i}^\alpha$ following a Gaussian distribution $f_{\beta_i}^\alpha\sim\mathcal{N}(1,\sigma_{\beta_i}^\alpha)$. For illustration purposes, we consider $\sigma_{\beta_i}^\alpha=0.15$. For the bid size (i.e., the offered flexibility quantities), we follow a similar approach to scale up/down the quantity of each bid, and we choose $\sigma_{\beta_i}^\alpha=0.3$. 
%
%
%
%
This analysis allows us to compare the statistical moments of the outputs of the LP and SOCP formulations. The numerically computed mean and the coefficient of variation (CV) -- i.e., the ratio of the standard deviation to the mean -- of the DLMPs and apparent power flows 
of the 141-bus system are shown in Fig.~\ref{fig:DLMP} and Fig.~\ref{fig:ApparentFlow}, respectively. The reported results correspond to simulations using 1000 Monte Carlo samples, as the numerical mean and CV converge after 1000 iterations.\vspace{-0.4cm}

\vspace{-0.4cm}

As shown in Fig.~\ref{fig:DLMP}, the averages of the DLMPs generated from the LP formulation closely reflect those of the SOCP formulation with small variations (a maximum variation of $1.85\%$ occurs at the root node). The CV of the DLMPs using both formulations is minimal showing close similarity of results. By contrast, even through the averages of the apparent power flows from the LP formulation closely approximate those from the SOCP formulation, as shown in Fig.~\ref{fig:ApparentFlow}, The CVs are more distinct compared with the results of DLMPs. As shown in Fig.~\ref{fig:ApparentFlow}, most of the CVs resulting from the LP formulation closely reflect those under SOCP with a few exceptions. For example, the CV of line number 86 in the LP formulation exceeds 300\% while that of the SOCP formulation stays below 200\%. This indicates that, in some samples, the apparent power flows over some lines in the LP formulation may show more variations than those in the SOCP formulation. Nevertheless, this observation does not undermine the LP formulation for congestion management because, as observed in our results, the larger CVs have appeared over lines with relatively small apparent power flows; hence, at low risk of being congested.  

\section{Conclusion and Future Research Outlook}
This paper has compared the results of an LP congestion management flexibility market formulation with those obtained under an SOCP model. The comparison has considered deterministic and Monte Carlo statistical case studies using two test systems. The obtained results have demonstrated the potential of the LP formulation in reliably capturing the results of its SOCP counterpart and that i) an increasing SL of flexibility improves the LP approximation, and ii) relaxation of voltage limits can improve the quality of the LP DLMPs and power flows at the cost of a lower-quality voltage approximation. The derived results are system-specific. Hence, for practical implementations (such as in the demonstration campaigns of the H2020 CoordiNet project), our presented analyses can be replicated to capture the underlying distribution system. In general, devising generalized results would require consideration of additional test systems and a dedicated mathematical comparison, which constitute key next research steps.

\section{Acknowledgements}
This work is supported by the European Union's Horizon 2020 research and innovation programme under grant agreement No 824414 -- CoordiNet Project.

\section{References}
\bibliographystyle{IEEEtran}
\bibliography{reference}

\end{document}